\documentclass[fleqn,twoside]{article}
\usepackage{espcrc2}

\usepackage{epsfig}
\usepackage[figuresright]{rotating}

\newcommand{\AmS}{{\protect\the\textfont2
  A\kern-.1667em\lower.5ex\hbox{M}\kern-.125emS}}
\newcommand{\bes}{\begin{eqnarray}}
\newcommand{\ees}{\end{eqnarray}}
\newcommand{\bea}{\begin{array}}
\newcommand{\ea}{\end{array}}
\newcommand{\eq}[1]{eq.~(\ref{#1})}
\newcommand{\fig}[1]{Fig.~\ref{#1}}
\def\gbar{\overline{g}}
\def\mbar{\overline{m}}
\newcommand{\Ffrac}[2]{\frac{\displaystyle{#1}}{\displaystyle{#2}}}
\def\Lmax{L_{\rm max}}
\def\mumin{\mu_{\rm min}}
\def\sigp{\sigma_{\rm P}}
\def\psibar{\overline{\psi}}
\def\rmA{{\rm A}}
\def\rmR{{\rm R}}
\def\rmP{{\rm P}}
\def\ZP{Z_{\rmP}}
\def\ZA{Z_{\rmA}}
\def\pd{\partial}
\def\dd{{\rm d}}

\title{
{
\vspace{-3.0cm} \normalsize \hfill
\parbox{30mm}{HU-EP-02/36\\DESY 02-125\\September 2002}
}\\[15mm]
       Running quark mass in two flavor QCD
       \thanks{Talk presented by F. Knechtli under the original title
       ``Towards the computation of the strange quark mass with
       dynamical quarks''. Work supported in part by the 
       European Community's Human Potential Programme under contract
       HPRN-CT-2000-00145, Hadrons/Lattice QCD.}
       }

\author{F. Knechtli\address[HU]{Institut f{\"u}r Physik, 
        Humboldt-Universit{\"a}t zu Berlin, Invalidenstr. 110, 
        10115 Berlin, Germany},
        M. Della Morte\address[DESY]{DESY Zeuthen, Platanenallee 6, 15738 Zeuthen, Germany},
        J. Rolf\addressmark[HU], R. Sommer\addressmark[DESY], 
        I. Wetzorke\address[DNIC]{NIC/DESY Zeuthen, Platanenallee 6, 15738 Zeuthen, Germany},
        U. Wolff\addressmark[HU] (ALPHA collaboration)
       }

\begin{document}

\begin{abstract}
We present first results for the step scaling function $\sigma_{\rm P}$ of
the renormalization factor $Z_{\rm P}$ of the pseudoscalar density.
The simulations are performed within the framework of the Schr{\"o}dinger
functional with two flavors of O($a$) improved Wilson fermions.
The knowledge of $\sigma_{\rm P}$ is required to compute the
renormalization group invariant quark masses. We also study the performance
of a variant of the HMC algorithm using two pseudofermion fields.
\vspace{1pc}
\end{abstract}

\maketitle

\section{THE RGI MASS}

Lattice QCD is a theory which has as free parameters the bare gauge coupling $g_0$ and the
bare current quark masses $m_i\,,\;i=u,d,s,...$. The hadronic scales like $F_\pi$ or
$m_{\rm K}$ are connected to the perturbative high energy regime of QCD via the running of
renormalized couplings
\bes
 \gbar^2(\mu) = Z_gg_0^2\,, & & 
 \mbar_i(\mu) = Z_mm_i\,,
\ees
where $\mu$ is the renormalization scale. At high energies the $\Lambda$ parameter and the
renormalization group invariant (RGI) quark masses $M_i$ can be determined.

In the following we assume that the renormalization conditions are independent of the
quark masses themselves. The running of the renormalized couplings is described by the
renormalization group equations (RGE)
\bes
\mu\frac{\dd\gbar}{\dd\mu}=\beta(\gbar)\,, &&
\mu\frac{\dd\mbar_i}{\dd\mu}=\tau(\gbar)\mbar_i \,.
\ees
The $\beta$ and $\tau$ functions are non-perturbatively defined if this is true for
$\gbar$ and $\mbar_i$. Their perturbative expansions are
\bes
 \beta(\gbar) &
 _{\mbox{$\stackrel{\displaystyle\sim}{\scriptstyle \gbar\rightarrow0}$}} &
 -\gbar^3\{b_0+b_1\gbar^2+b_2\gbar^4+...\} \,, \\
 \tau(\gbar) &
 _{\mbox{$\stackrel{\displaystyle\sim}{\scriptstyle \gbar\rightarrow0}$}} &
 -\gbar^2\{d_0+d_1\gbar^2+...\} \,.
\ees
RGI quantities $P$ do not depend on the renormalization scale $\mu$
\bes
 \mu\frac{\dd}{\dd\mu}P(\mu,\gbar,\mbar_i) & = & 0 \,.
\ees
Examples are the $\Lambda$ parameter and the RGI quark masses
\bes
 \Lambda & = & \mu(b_0\gbar^2)^{-b_1/2b_0^2}
 \exp\left\{-\Ffrac{1}{2b_0\gbar^2}\right\}
 \label{LAMBDA} \\
 & & \times\exp\left\{-\int_0^{\gbar}\dd x\left[
 \Ffrac{1}{\beta(x)}+\Ffrac{1}{b_0x^3}-\Ffrac{b_1}{b_0^2x}\right]\right\}
 \nonumber
\ees
\bes
 M_i & = & \mbar_i(2b_0\gbar^2)^{-d_0/2b_0} \label{MRGI} \\
 & & \times\exp\left\{-\int_0^{\gbar}\dd x\left[
 \Ffrac{\tau(x)}{\beta(x)}-\Ffrac{d_0}{b_0x}\right]\right\} \,, \nonumber
\ees
where $\gbar=\gbar(\mu)$ and $\mbar_i=\mbar_i(\mu)$. These RGI parameters of QCD
are defined beyond perturbation theory and their connections between different
renormalization schemes can be given in a simple and exact way.
For these reasons $\Lambda$ and $M_i$ should be taken as the fundamental parameters of QCD.

In this talk we present first results on the non-perturbative computation of the
running of $\mbar(\mu)$ in the Schr{\"o}dinger Functional (SF) renormalization scheme
with two dynamical flavors of $O(a)$ improved Wilson fermions. The strategy of our
computation closely follows ref. \cite{Capitani,quenched}. We will give as a result
an approximation to the flavor independent ratio $M/\mbar(\mu)$
as a function of $\mu/\Lambda$.


\section{THE RUNNING MASS}

In continuum QCD a renormalized mass in the SF scheme is defined through the
PCAC relation which involves the renormalized axial current $(A_{\rmR})_\mu(x)$
and the renormalized pseudoscalar density $P_{\rmR}(x)$
\bes
 \pd_{\mu}(A_{\rmR})_{\mu} & = & (\mbar_i+\mbar_j)P_{\rmR} \\
 (A_{\rmR})_{\mu}(x) & = & \ZA\psibar_i(x)\gamma_{\mu}\gamma_5\psi_j(x) \\
 P_{\rmR}(x) & = & \ZP\psibar_i(x)\gamma_5\psi_j(x) \,.
\ees
The renormalization constant $\ZA$ is fixed by Ward identities. The renormalization
constant $\ZP$ is defined as
\bes
 \ZP(L) = \Ffrac{\sqrt{3f_1}}{f_{\rmP}(L/2)} \; \mbox{at} \;
 m_i=0\,,\;i=u,... \,.
\ees
For notation and details of the calculation we refer to \cite{Capitani,Sint}. 
The correlation functions $f_1$ and $f_P$ are computed at zero bare current quark masses
$m_i$. The renormalization scale in the SF scheme is identified with the inverse spatial box size
$\mu=1/L$. Keeping the physical size $L$ constant can be realized by keeping the
renormalized coupling $\gbar(L)$ fixed.

A rigorous definition of the renormalized mass can be given in the lattice
regularization of QCD by
\bes
 \mbar_i(\mu) = \left.\lim_{a\to0}\,
                \Ffrac{\ZA(g_0)}{\ZP(g_0,L/a)}m_i(g_0)\right|_{u=\gbar^2(L)} \,. &&
\ees
The running of the mass can be described by the flavor independent step scaling
function $\sigp(u)$
\bes
 \frac{\mbar(\mu)}{\mbar(\mu/2)}=
 \left.\lim_{a\to0}\,\frac{\ZP(g_0,2L/a)}{\ZP(g_0,L/a)}\right|_{u=\gbar^2(L)}
 = \sigp(u) \,. &&
\ees
Together with the step scaling function for the gauge coupling
$\sigma(u)=\gbar^2(2L)$ \cite{Michele} we can solve a system of coupled
recursions: $k=0,1,2,...$
\bes
 & & \left\{\bea{l} u_0=\gbar^2(\Lmax)=3.3000 \\ 
 \sigma(u_{k+1})=u_k \ea\right. \label{couplrec}
 \\
 & & \Rightarrow u_k=\gbar^2(2^{-k}\Lmax) \nonumber
\ees
\bes
 & & \left\{\bea{l} v_0=1 \\ 
 v_{k+1}=\Ffrac{v_k}{\sigp(u_k)} \ea\right. \label{massrec}
 \\
 & & \Rightarrow v_k=\Ffrac{\mbar(\mumin)}{\mbar(2^k\mumin)} 
     \; \mbox{at} \; \mumin=\frac{1}{2\Lmax} \,. \nonumber
\ees
The largest coupling $u_0$ defines the reference scale $\Lmax$. At high energies
contact with perturbation theory can safely be made and the RGI
parameters are determined from \eq{LAMBDA} and \eq{MRGI} using
the perturbative 2--loop $\tau$ function and 3--loop $\beta$ function.
%
\begin{figure}[t]
\vspace{-1.7cm}
\centerline{\epsfig{file=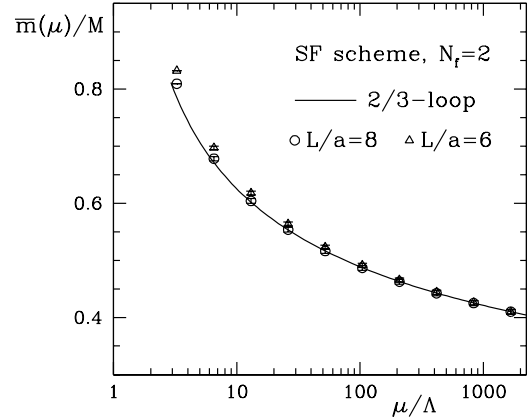,width=9cm}}
\vspace{-2.0cm}
\caption{The running of the renormalized quark mass. \label{f_running}} 
\end{figure}
%

We computed the lattice step scaling function $\Sigma_{\rm P}(u,a/L)$ for 6 couplings
in the range $u=3.33\;...\;0.98$ on $L/a=6$ and $L/a=8$ (requiring also $2L/a$) lattices.
We interpolate $\Sigma_{\rm P}$ by the Ansatz
\bes
 \Sigma_{\rm P}(u,a/L) & = & 1 -\ln(2)d_0u + p_2u^2 + p_3u^3
\ees
with fitted parameters $p_2$ and $p_3$.
Using it in the recursion \eq{massrec} we obtain
the following approximations for
\bes
 \frac{M}{\mbar(\mumin)} & : & \left\{
 \bea{l} {1.202(13) \quad L/a=6} \\ 
         {1.236(15) \quad L/a=8} \ea\right.
\ees
(contact with perturbation theory is made at the scale $2^{-8}\Lmax$).
Combining with recursion \eq{couplrec}, from which we get in the continuum limit
$\ln(\Lambda\Lmax)=-1.85(13)$ \cite{Michele}, we plot in \fig{f_running} the points
\bes
 \frac{\mbar(\mu_k)}{M} & = & \frac{1}{v_k}\frac{\mbar(\mumin)}{M}
\ees
as a function of $\mu_k=2^k\mumin$, together with the pertubative curve using the
2--loop $\tau$ function and the 3--loop $\beta$ function. 
The errors of the points come from the statistical errors in the coefficients $v_k$.
The overall uncertainties in the quantities $\Lambda\Lmax$ and $M/\mbar(\mumin)$ are not
shown. The points show a dependence on the lattice spacing but for an extrapolation
to the continuum limit we need additional simulations on finer lattices.
%
\begin{figure}[t]
\vspace{-0.1cm}
\centerline{\epsfig{file=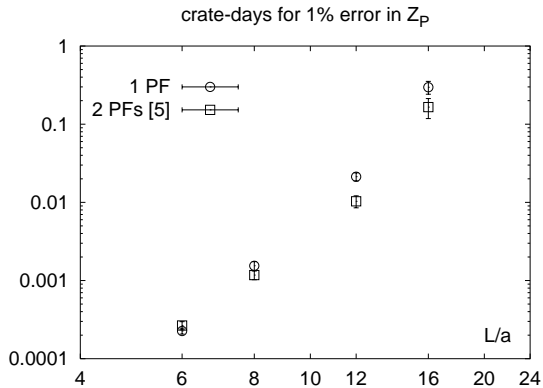,height=7.8cm,angle=-90}}
\vspace{-0.8cm}
\caption{Performance of the HMC algorithm using one or two psedofermion fields (PFs). 
         \label{f_cost}}
\end{figure}
%

\section{ALGORITHM STUDY}

In our simulations we compared the computational cost of the HMC algorithm 
for the standard action and for an action 
which uses two pseudofermion fields (PFs) \cite{Martin}.
We denote by $\hat{Q}$ the even--odd preconditioned hermitian Dirac operator for
$O(a)$ improved Wilson fermions. 
The effect of $\det(\hat{Q}^2)$ is taken into account by two PFs $\phi_1$ and $\phi_2$
with actions
\bes
 S_{{\rm F}_1}& = & 
 \phi_1^{\dagger}\left[(\hat{Q}^2+\rho^2)^{-1}\right]\phi_1 \,, \label{sb1} \\
 S_{{\rm F}_2} & = &
 \phi_2^{\dagger}\left[\rho^{-2}+\hat{Q}^{-2}\right]\phi_2 \,. \label{sb2}
\ees
This splitting of the action is valid for any real parameter $\rho$.
We take
\bes
 \rho & = &
 \left(\lambda_{\min}(\hat{Q}^2)\lambda_{\max}(\hat{Q}^2)\right)^{1/4} \,,
\ees
which minimizes the sum of the condition numbers in \eq{sb1} and \eq{sb2}.
This is expected to reduce the fermionic force allowing for larger step sizes in
molecular dynamics.

The implementation of the Hybrid Monte Carlo (HMC) algorithm to simulate
the system described by \eq{sb1} and \eq{sb2} is straightforward.
Simulations with the same physical parameters show that
the step size of molecular dynamics with two PFs
can be doubled with respect to the standard case with one PF
by keeping the same acceptance (optimal around 80\%). In \fig{f_cost} we show
the performance of the HMC algorithm on the APEmille machine, 1 crate of the machine
consists of 128 nodes and sustains 68 GFlops peak. The computational cost to achieve
1\% error in $\ZP$ at constant value of $\gbar^2\approx2.48$
is almost a factor 2 lower for two PFs than for one PF on lattices
$L/a=12$ and $L/a=16$. This gain in performance is important in view of the next
simulations on lattices $L/a=12$ and $2L/a=24$ needed to obtain $\sigp(u)$ in
the continuum limit.

{\bf Acknowledgement.} We thank NIC/DESY Zeuthen for allocating computer time on the
APEmille machine indispensable to this project and the APE group for their support.

\end{document}